\documentclass[
    ,final            
  ]
  {aipproc}

\layoutstyle{6x9}

\begin{document}

\title{What is the origin of the soft excess in AGN?}


\author{Ma{\l}gorzata Sobolewska}{
  address={Copernicus Astronomical Center, Warsaw, Poland}
}

\author{Chris Done}{
  address={University of Durham, Durham, UK}
}

\begin{abstract}
We investigate the nature of the soft excess below 1~keV observed in
AGN. We use the XMM-Newton data of the low redshift, optically bright
quasar, PG~1211+143, and we compare it with the Narrow Line Seyfert 1
galaxy, 1H~0707-495, which has one of the strongest soft excesses
seen. We test various ideas for the origin of the soft X-ray excess,
including a separate spectral component (for example a low temperature
Comptonized component), a reflection-dominated model, or a complex
absorption model. All three can give good fits to the data, and
$\chi^2$ fitting criteria are not sufficient to discriminate among
them. Instead, we favor the complex absorption model on the grounds
that it is the most physically plausible.
\end{abstract}

\maketitle


\section{Introduction}

Several models are considered in the literature to explain the origin
of the soft excess below 1~keV in AGN, which cannot be explained by
either of the two emission components usually considered in accreting
black holes, namely the accretion disk and the high energy power law
Comptonized emission. Spectra with a strong soft excess also often
show a strong deficit at $\sim 7$~keV which again has no obvious
identification. 

{\bf Partial covering models} allow for a spatially non-uniform cold
absorber attenuating only a fraction of intrinsic radiation. They can
explain the sharp features around 7~keV. However, they require an
enormous overabundance of iron (5-30 times the Solar value
\cite{Tanakaea:2004}; 5 times the Solar value \cite{Galloea:2004}),
and they cannot simultaneously explain the soft X-ray emission. Thus,
we do not consider partial covering models in our analysis as they
only provide a method to explain the sharp feature around 7~keV. 

The soft excess may originate in a distinct unknown physical process
which manifests itself as {\bf a separate spectral component}
(Model~1). It can be modeled e.g. with low temperature Comptonization
models \cite{Magdziarzea:1998}). The apparent temperature of this
component is rather similar ($\sim 0.1-0.3$~keV) across a diverse
selection of AGN (e.g. \cite{Czernyea:2003}). 

An obvious way to produce a fixed soft excess energy is to relate it
to atomic processes rather than continuum emission. In reflection
models, the dramatic strength of the soft excess requires that the
source is {\bf reflection  dominated} (Model~2). This might happen if
the disc fragments at high accretion rates, hiding the hard X-ray
source among many clumps. Such models can successfully fit the
spectra, but need a large range in ionization states of the reflecting
material, and supersolar abundances (\cite{Fabianea:2002}).  

Alternatively, the strong jump in opacity for partially ionized
material could result in a soft excess from {\bf absorption} 
(Model~3, \cite{GierlinskiDone:2004a}). The underlying continuum may
be modeled by only one component.  The soft excess and hardening of
the spectrum at high energies would be only artifacts of absorption
acting at 1-2~keV, while the H-like iron K alpha resonance
absorption from the same material could produce the deficit at $\sim
7$~keV.   

\section{Results}

The best fit models (and their components) together with the data to
model ratios are shown in Figure~1. The models are additionally
modified by galactic absorption, cold absorption at the redshift of
the source, and one (in 1H~0707-495) or two (in PG~1211+143) warm
absorbers modeling the narrow absorption features. Thus, the only
thing that differentiates the three models is the description of the
origin of the soft excess.  

\begin{figure}[ht]
  \includegraphics[height=.2\textheight]{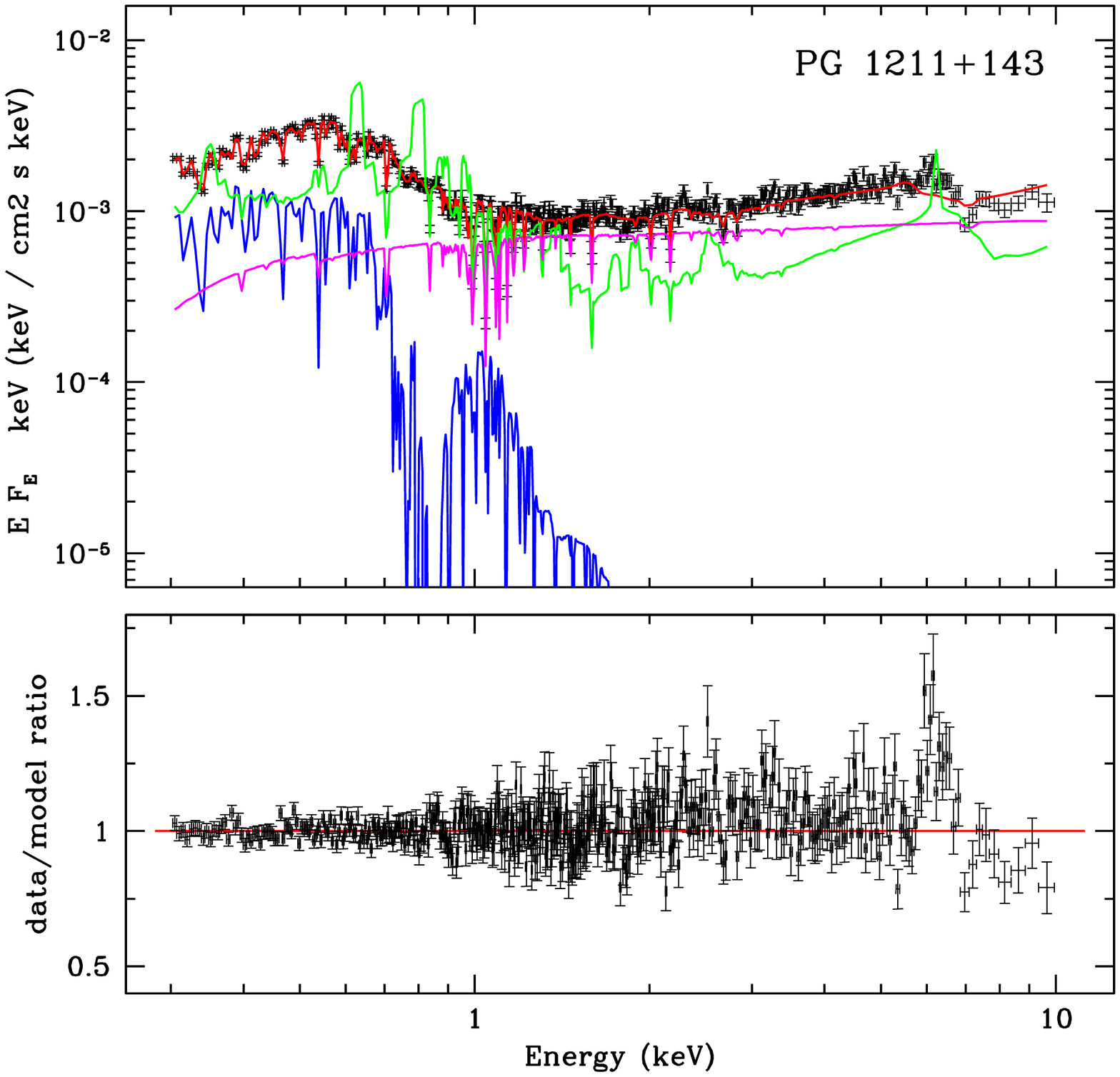}
  \includegraphics[height=.2\textheight]{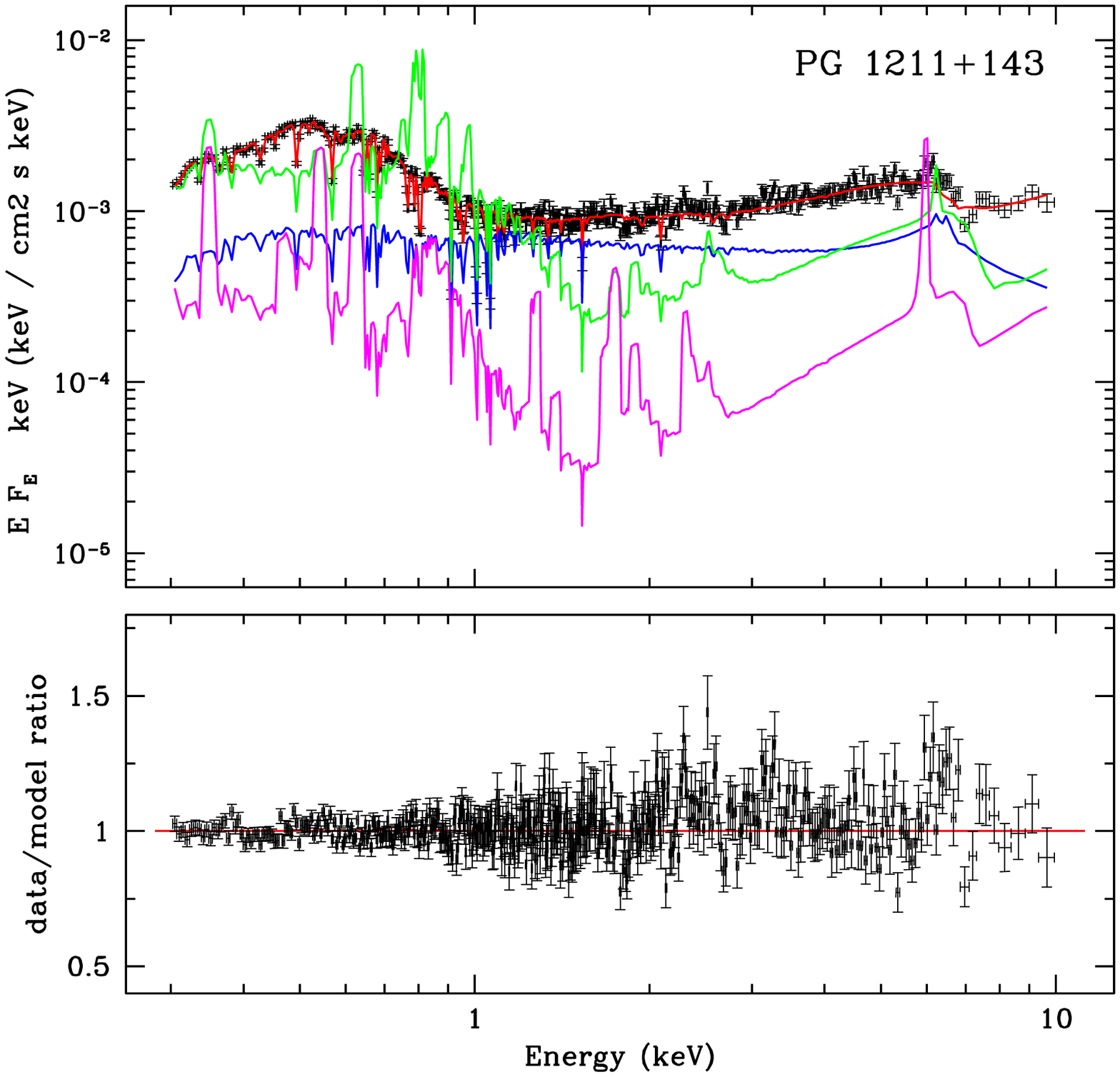}
  \includegraphics[height=.2\textheight]{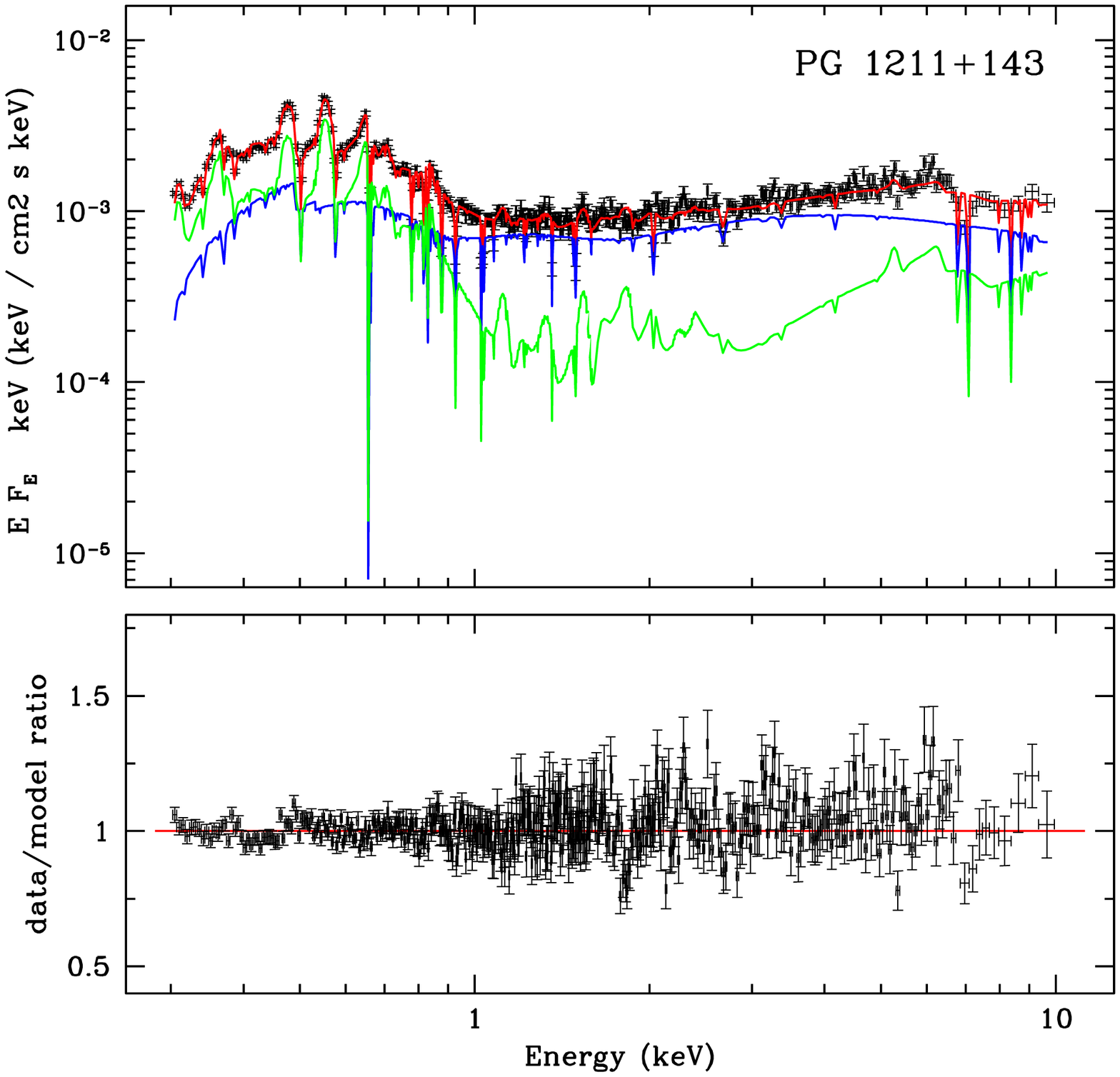}
\end{figure}
\begin{figure}[ht]
  \includegraphics[height=.2\textheight]{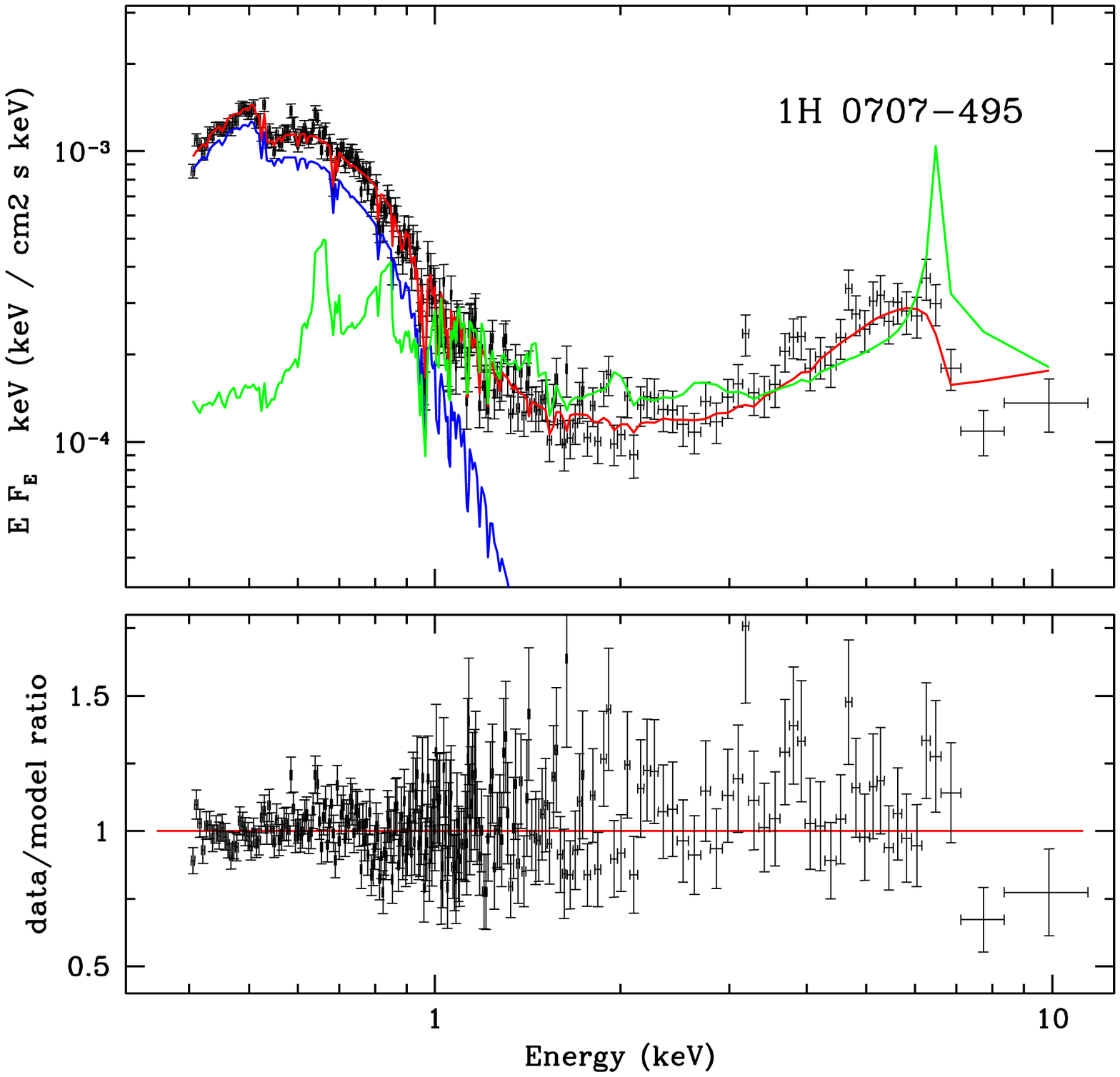}
  \includegraphics[height=.2\textheight]{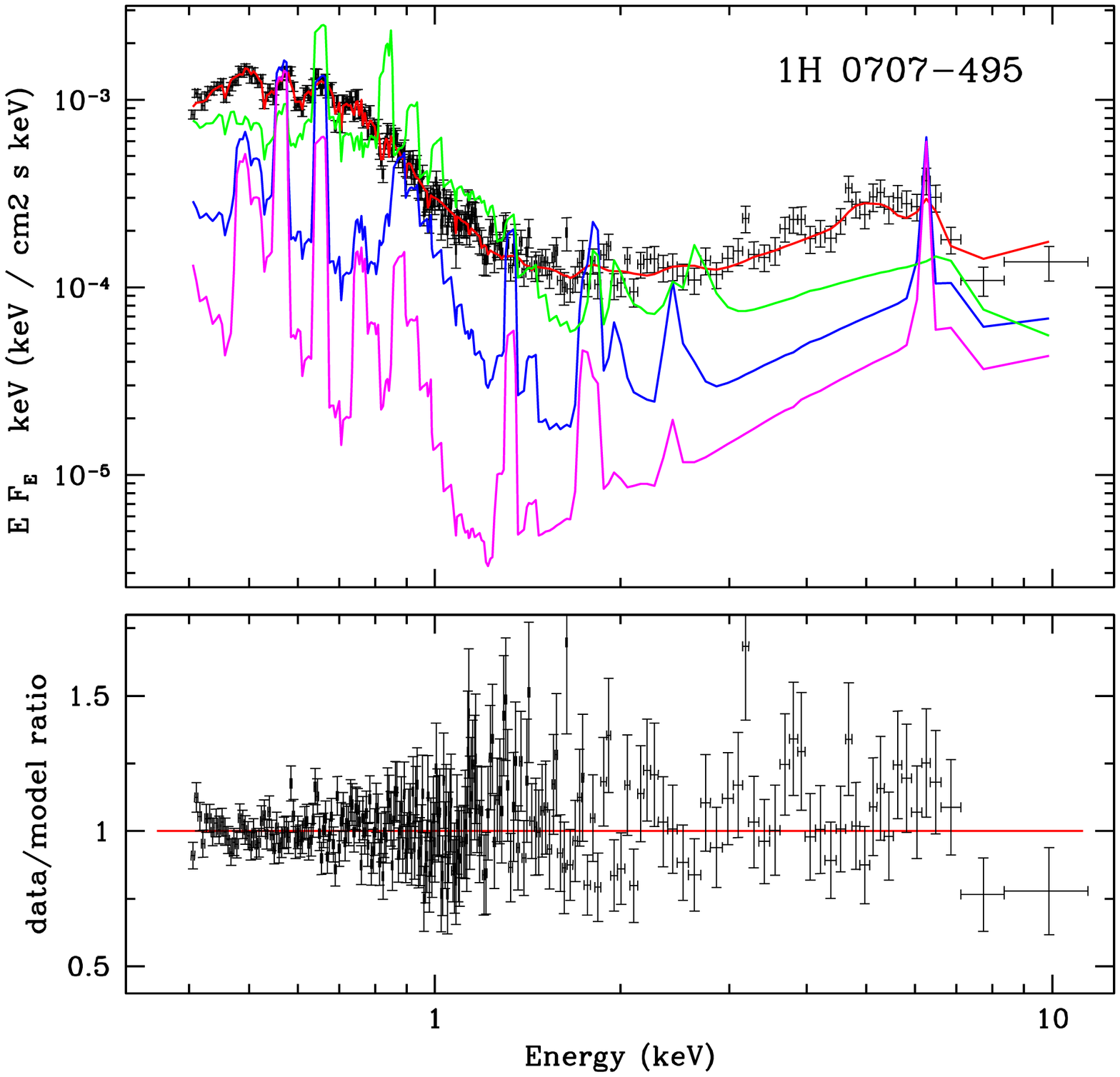}
  \includegraphics[height=.2\textheight]{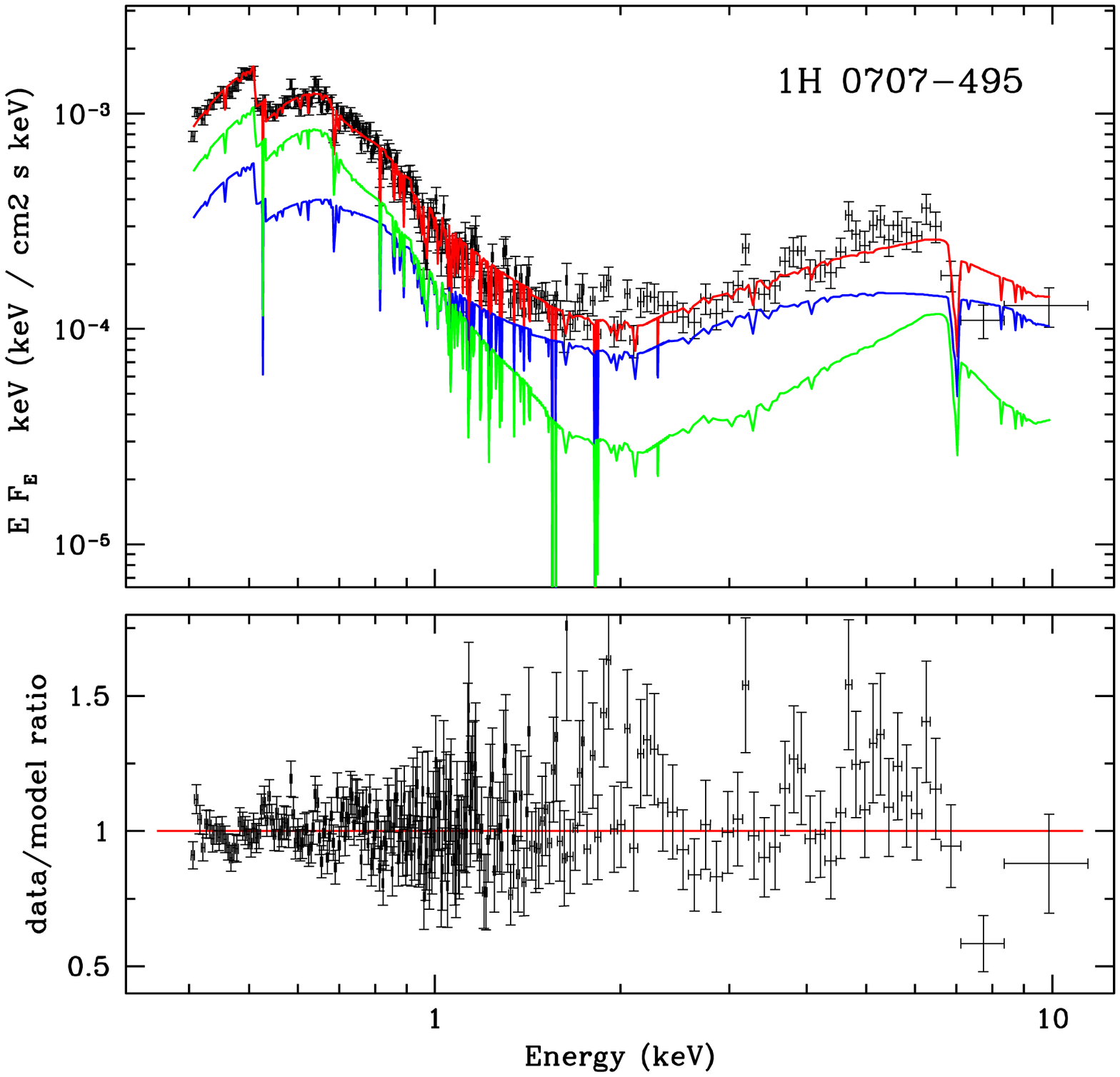}
  \caption{Modeling the soft excess in PG~1211+143 and
  1H~0707-495. {\it Red:} the best fit spectra. {\it Left:} Model~1 --
  a low temperature Comptonization ({\it blue}), power law ({\it
  magenta}) and ionized reflection ({\it green}). In 1H~0707-495 there
  is no need for the intrinsic power law. {\it Middle:} Model~2 --
  three reflectors ({\it green}, {\it magenta} and {\it blue}) with
  different column densities and ionization states, located at
  different radii. {\it Right:} Model~3 -- a power law ({\it blue})
  and reflection ({\it green}) subject to relativistically smeared
  absorption.}
\end{figure}

\begin{table}[ht]
\begin{tabular}{l c c c}
\hline
  \tablehead{1}{l}{b}{Object\\~}
  & \tablehead{1}{c}{b}{Model 1\\~}
  & \tablehead{1}{c}{b}{Model 2\\$\chi^2$ ({\it  d.o.f.})}
  & \tablehead{1}{c}{b}{Model 3\\~} \\
\hline
PG 1211+143 & 989 (939) & 972 (937) & 998 (939)\\
1H 0707-495 & 311 (288) & 292 (283) & 318 (287)\\
\hline
\end{tabular}
\caption{Results of spectral fits to the PG 1211+143 and 1H 0707-495 data}
\label{tab:a}
\end{table}

In PG~1211+143 the three models give virtually the same quality fits
in terms of the reduced $\chi^2 = 1.04$--1.06. In 1H~0707-495 the
model of complex absorption results in a fit comparable to that of the
model with a distinct spectral component (with reduced $\chi^2$ of
1.10 and 1.08, respectively). The best fit is obtained with the model
of complex reflection (with iron abundance set to 2 times the Solar
abundance) with reduced $\chi^2$ of 1.03. The values of $\chi^2$ and
number of degrees of freedom in fits are presented in Table~1.

\section{Discussion and conclusions}

Based only on the $\chi^2$ criterion it is not possible to uniquely
distinguish between the three kinds of models as each of them provides
an adequate description. Moreover, many {\bf model} uncertainties can
contribute to $\chi^2$, e.g. in both the reflection and absorption
scenarios we expect an (unknown!) range of ionization states to be
present. Hence direct comparison of $\chi^2$ can be misleading when
the models are known to be incomplete. Instead, we should be guided as
well by physical plausibility. 

The model with a distinct component involves an emission from an
unknown physical process, with unknown process fixing its typical
energy, which does not seem to have an analogue in the spectra of
galactic black holes (GBHs) \cite{GierlinskiDone:2004b}, and cannot
simultaneously explain the 7~keV feature.  

By contrast, both reflection and absorption are much more plausible,
as they give a physical reason fixed energy for the soft excess and
can reproduce the structure around iron K. However, the reflection
models require quite strong supersolar abundances, and there is no
evidence for reflection dominated spectra in the GBH systems.

The absorption model seems to reproduce the strong soft excesses well
with moderate column densities ($\sim 10^{22-23}$~cm$^{-2}$) of solar
abundance material. Such material would not be seen in the GBHs
because it would be completely ionized by the much higher accretion
disc temperatures seen in the stellar mass black hole systems. Thus
the complex absorption model provides an interesting alternative
because (1) it does not require a separate soft component in the
spectra, (2) the hard radiation slope resulting from the fits is
similar to that of GBHs in the high/soft state, (3) it describes the
spectra of AGNs in terms of atomic processes, explaining the universal
shape of the soft excess.


\begin{theacknowledgments}
This work was partially supported by the Polish Committee for
Scientific Research through grants 1P03D01626 and 2P03D00322.
\end{theacknowledgments}


\end{document}